\documentclass[11pt,a4paper]{article}
\pdfoutput=1
\usepackage{jcappub}
\usepackage[T1]{fontenc} 

\usepackage{graphicx}
\usepackage{dcolumn}
\usepackage{bm}
\usepackage{hyperref}
\usepackage{soul}

\usepackage{graphicx}
\usepackage{dcolumn}
\usepackage{bm}
\usepackage{hyperref}
\def\eq#1 { \begin{equation} #1 \end{equation} }
\def\eqn#1{ \begin{eqnarray} #1 \end{eqnarray} }
\def\nn { \nonumber }
\def\Re{{\rm Re}\,}
\def\Im{{\rm Im}\,}

\begin{document}

\title{A local Lagrangian for MOND as modified inertia}

\author[1,2]{Renato Costa,}
\author[1,2,3]{Guilherme Franzmann} 
\author[]{and}
\author[4,5,6]{Jonas P. Pereira}

\affiliation[1]{The Cosmology and Gravity Group, Department of Mathematics and Applied Mathematics, University of Cape Town, Private Bag, Rondebosch, 7700, South Africa}
\affiliation[2]{Physics Department, McGill University, Montreal, QC, H3A 2T8, Canada}
\affiliation[3]{Center for Gravitational Physics, Yukawa Institute for Theoretical Physics, Kyoto University, 606-8502, Kyoto, Japan}
\affiliation[4]{Universidade Federal do ABC, Centro de Ci\^encias Naturais e Humanas, Avenida dos Estados 5001- Bang\'u, CEP 09210-580, Santo Andr\'e, SP, Brazil}
\affiliation[5]{Mathematical Sciences and STAG Research Centre,
University of Southampton, Southampton, SO17 1BJ, United Kingdom}
\affiliation[6]{Nicolaus Copernicus Astronomical Center, Polish Academy of Sciences, Bartycka 18, 00-716, Warsaw, Poland}
\emailAdd{renato.costa87@gmail.com}
\emailAdd{guilherme.franzmann@mail.mcgill.ca}
\emailAdd{jpereira@camk.edu.pl}

\abstract{We propose a local Lagrangian for a point particle where its inertia part is modified in the regime of small accelerations. 
For the standard gravitational central force, it recovers the deep MOdified Newtonian Dynamics (MOND) (accelerations $\ll a_0\approx 10^{-10}$m\,s$^{-2}$) equations of motion in the case of a circular orbit.
Perturbations to that turn on higher derivative terms, leading to exponentially unstable solutions that must vanish in order to account for the very small scattering of the Tully-Fisher relation. Unstable solutions linearly growing with time remain valid for a characteristic timescale of at least 3 billion years.
We show that vertical perturbations recover similar results to dark matter for old galaxies, but deviations could be present for young ones.  We also present ways to probe our approach and describe some of its subtleties, such as the strong equivalence principle (violated in general), the center of mass motion of a composite body, and how in some cases it could overcome Ostrogradsky's instabilities (with naturally occurring piecewise Lagrangians). Our main conclusions regarding our MOND-like proposal are: (i) it constitutes a possible recipe where Ostrogradsky instabilities could be ``tamed''; (ii) it is a falsifiable approach in various contexts and (iii) it might explain simultaneously some of the issues usual modified gravity MOND and dark matter phenomenologies have difficulties individually. These aspects seem relevant to start addressing practical ways to differentiate modified gravity MOND from modified inertia and give insights into alternative ways to tackle some astrophysical and cosmological puzzles.}


\maketitle

\flushbottom

\section{Introduction}

Observations of orbital velocities of stars have shown that when Kepler's laws are kept the dynamical masses of spiral galaxies should be much larger than their baryonic masses. This is the cradle of dark matter, a nonluminous form of matter that should mainly interact gravitationally with ordinary matter and be present in the outer regions of galaxies. Dark matter's (DM) importance has gone much beyond local aspects of the universe and now is one of the cornerstones of the $\Lambda$CDM cosmological model, playing a fundamental role in structure formation and fitting the CMB power spectrum (for a review see \cite{Garrett:2010hd}). Yet, it faces some challenges in galactic scales. The observations point towards strong correlation between asymptotic velocities in the outskirts of different types of galaxies and their baryonic masses, which can be fit invoking a single parameter: an acceleration scale of order $a_0 \sim 10^{-10}$ ms$^{-2}$. A particular case corresponds to the Baryonic Tully-Fisher relation, which provides the scaling for spiral galaxies. Although it is possible that dark matter can recover such relations, simulations seem to imply a much bigger scattering than what is observed \cite{Lelli:2015wst} (see \cite{Famaey:2011kh,Famaey:2013ty} for a more complete list of challenges to the dark matter paradigm).

There have been many proposals to explain such scaling relations \citep{Lelli:2017vgz},
however Milgrom's MOdified Newtonian Dynamics (MOND) \cite{Milgrom:1983ca,Sanders:2002pf,Famaey:2011kh} is probably the most popular in this regard. It is a phenomenological model conceived in the context of rotation curves which for instance naturally predicts the Tully-Fisher relation  \cite{Tully:1977fu,1983ApJ...270..371M,2005ApJ...632..859M}.
Currently, MOND in astrophysics is mostly understood as a modification of the gravitational force \cite{Bekenstein:1984tv}. As such, it has as one consequence the nonlinearity of the gravitational field equations, which for instance results in the violation of the strong equivalence principle (SEP). SEP's violation is known in the jargon of MOND as the ``External Field Effect'' (EFE). It has highly nonintuitive consequences, for instance the ``Newtonization'' of any system under a strong gravitational field ($|\vec{g}|\gg a_0 \approx 10^{-10}$ m\,s$^{-2}
$) \cite{Bekenstein:1984tv}.
It can also be used to explain why Earth-based experiments so far have not detected any deviation from Newton's dynamics as well as how test particles can escape from a gravitational field \cite{Gundlach:2007zz,Famaey:2007tc}.

MOND could also be extended to other interactions via the modified gravity recipe. Similarly to the gravitational case, the price to pay would be the modification of the underlying field equations for the fundamental interactions. In general, the field equations would become nonlinear.
Nevertheless, one could also interpret MOND on the level of rotation curves as a modification of Newton's second law, keeping fundamental fields unchanged. One undesirable consequence of that would be the need of a nonlocal theory for test particles (a Lagrangian with an infinite number of time derivatives) when the usual Milgrom's law of motion is imposed \cite{Milgrom:1992hr,Milgrom:2005mc,Milgrom:2011kx}.

In this work we show that there is a simple solution to nonlocality in MOND's modified inertia by means of a generalized law of motion (higher than second order, though) which has MOND as a particular case. More specifically, we show that there always exists a local Lagrangian whose equations of motion lead to usual MOND for circular orbits in spiral galaxies, but differs in other cases. We show that this Lagrangian is unstable under small perturbations in the deep MOND regime\footnote{In our approach, as in usual MOND, we assume that the deep MOND regime is defined by acceleration magnitudes $(|\vec{a}|)$ much smaller than $a_0$. The Newtonian regime in our proposal would also be attained when $|\vec{a}|\gg a_0$. However, whenever ambiguities might rise, we call our approach ``MOND-like''.}, but it might present stability 
for interactions other than gravity.
However, subtleties are present and some classical concepts such as the SEP and the invariance of the equations of motion under Galilean transformations break down when the total acceleration $|\vec{a}|$ is much smaller than $a_0$ (we elaborate on these issues later on, in our discussion section.)
One advantage of our analysis is that with a local Lagrangian at hands, it becomes trivial the issue of the equations of motion in noninertial frames, a crucial step for tests on Earth of versions of MOND.

Our analysis is also motivated by some consequences of the recent direct detection of gravitational waves from a binary neutron star inspiral by the LIGO/VIRGO collaboration \citep{2017PhRvL.119p1101A}.
The reason is as follows. It has been found as a consequence of the 1.7s delay between the gravitational wave and electromagnetic detections that the speed of gravitational waves could differ at most (when conservative assumptions are taken into account) by one part in $10^{15}$ from the speed of light in the vacuum \citep{2017ApJ...848L..13A}.
This severely constrains alternative theories to general relativity (see, e.g.,  \citep{2017PhRvL.119y1304E,Sanders:2018jiv,Sakstein:2017xjx,2019CQGra..36a7001C} and references therein), even some candidates for a relativistic completion of Milgrom's phenomenological MOND \citep{2017PhRvL.119c1102C}. At the same time, it points to a universality of chargeless test particle trajectories, given that their velocities would differ negligibly. Since we basically deal with new equations of motion and keep fields unaltered, we thus have that our approach is in qualitative agreement with LIGO measurements.

This work is structured as follows: In Section \ref{1Dmond_setup} we present the general one-dimensional approach to obtain a Lagrangian that includes a MOND-like term and use it to test the dynamics of a Hooke-like force. Since this toy model leads to a higher order differential equation, we also comment about the Ostrogradsky theorem and how it is circumvented in this case. Section \ref{MOND3D} extends the toy model to central potentials and we show that in the particular case of a circular motion in a gravitational central force, MOND's equations of motion are obtained. In section \ref{MOND2D} we analyze the two dimensional case and explicitly show that the deep MOND solution is unstable under small radial and vertical perturbations around circular orbits, but these instabilities are either phenomenologically discarded or used to define the regime of validity of perturbation theory, found to be at the very least of order of some billions of years. 
Besides, we investigate an observationally motivated interpolating function and find that for the intermediate MOND regime instability timescales would be smaller than a billion years.
In section \ref{CM}, we start addressing the nontrivial issue concerning the center of mass motion in our theory of local modified inertia, and show a specific example where the internal and external dynamics can be decoupled in a system of multiple bodies. Finally,
section \ref{conclusions} contains our discussions and conclusions.

\section{1D MOND-like in the deep MOND regime}
\label{1Dmond_setup}

We start by recalling that the standard Newtonian equations of motion for a point particle can be derived from the Euler-Lagrange equations by the Lagrangian
\eqn{
L(\dot{q},q) = \frac{m}{2}\dot{q}^2 - U(q),
}
or equivalently by the generalized Euler-Lagrange equations by the Lagrangian
\eqn{
L(a,\dot{q},q) = -\frac{m}{2}\vec{a}\cdot \vec{q} - U(q).\label{Milg}
}

Milgrom's dynamics, however, introduces an acceleration scale by proposing an equation of motion quadratic in the acceleration for a certain acceleration regime. A natural question to ask is whether there is a generalization of Lagrangian (\ref{Milg}) that recovers Milgrom's dynamics. In order to do so, consider the most general Lagrangian that depends on the generalized acceleration, $a$, velocity, $\dot{q}$, and position, $q$, in one dimension,  
\eqn{
L = L(a,\dot{q},q).
}
From the generalized Euler-Lagrange equation we get
\eqn{\label{general_equation}
&-&\dot{a}\bigg[\frac{\partial^3 L}{\partial a^3} \dot{a}+\frac{\partial^3 L}{\partial a^2 \partial \dot{q}} a+\frac{\partial^3 L}{\partial a^2 \partial q} \dot{q}\bigg] - \frac{\partial^2 L}{\partial a^2} \ddot{a}
-a\bigg[\frac{\partial^3 L}{\partial a^2\partial \dot{q}} \dot{a}+\frac{\partial^3 L}{\partial a \partial \dot{q}^2} a+\frac{\partial^3 L}{\partial a \partial \dot{q} \partial q} \dot{q}\bigg] - \frac{\partial^2 L}{\partial a \partial \dot{q}} \dot{a}
\nonumber\\
&-&\dot{q}\bigg[\frac{\partial^3 L}{\partial a^2 \partial q} \dot{a}+\frac{\partial^3 L}{\partial a \partial \dot{q} \partial q} a+\frac{\partial^3 L}{\partial a \partial q^2} \dot{q}\bigg] - \frac{\partial^2 L}{\partial a \partial q} a
+\frac{\partial^2 L}{\partial a \partial \dot{q}} \dot{a}+\frac{\partial^2 L}{ \partial \dot{q}^2} a+\frac{\partial^2 L}{\partial \dot{q} \partial q} \dot{q} - \frac{\partial L}{\partial q}=0.}
The only terms that can give rise to the pure quadratic term in acceleration, typical of the deep MOND regime, are:
\eqn{
-a^2 \frac{\partial^3 L}{\partial a \partial \dot{q}^2},\qquad - \frac{\partial^2 L}{\partial a \partial q} a,\qquad \frac{\partial^2 L}{\partial \dot{q}^2}a, \qquad -\frac{\partial L}{\partial q}, 
}
as long as
\eqn{
 \frac{\partial^3 L}{\partial a \partial \dot{q}^2} &=& c_1,
 \label{c1}
 \\
 \frac{\partial^2 L}{\partial a \partial q} &=&  c_2 a,
 \label{c2}
 \\ 
 \frac{\partial^2 L}{\partial \dot{q}^2} &=& - c_3 a
 \label{c3},
 \\
 \frac{\partial L}{\partial q} &=& 
 c_4 a^2\label{c4},
}
where $c_1, c_2, c_3$ and $c_4$ are real constants.
Integrating terms (\ref{c1}) and (\ref{c3}) leads to a total derivative and do not contribute to the equation of motion while integrating (\ref{c2}) and comparing to (\ref{c4}) implies $c_2=2c_4$. Thus, the Lagrangian is
\eqn{	
L = - 2c_4 a^2 (q-q_0) + f(\dot{q},q),
\label{MOND1D}}
where $q_0$ is a reference position naturally associated with a frame choice to define a physical distance $(q-q_0)$ to the problem and $f(\dot{q},q)$ is an arbitrary function of the position and velocity variables.
We will see in Section \ref{MOND3D} that the deep MOND regime for a gravitational force fixes $2c_4= m/3a_0$ and that Eq. \eqref{MOND1D} is the correct form for the Lagrangian in order to recover the Tully-Fisher relation. By choosing $f = U(q)$, we thus have that the simplest deep MOND-like Lagrangian is given by
\eqn{
L(a,q) = -\frac{m}{3} \frac{a}{a_0} \vec{a} \cdot  (\vec{q}-\vec{q_0}) - U(q),
\label{L3D_simplest}}
where $U(q)$ is an arbitrary function of $(q-q_0)$, which represents the potential energy, $a = |\vec{a}|$ and the inner product reinforces that the direction of acceleration is the same as the force under consideration. Note that this is valid for any kind of force, not only gravitational. Finally, we stress that if one assumed the Lagrangian was a function of higher order derivatives of the acceleration, then one would need an infinite number of them in order to recover the usual Milgrom's equation of motion. This is why it is related to a nonlocal theory. As a consequence, local theories will, in general, lead to equations of motion with higher order derivatives (different from Milgrom's).

With the use of the Euler-Lagrange equation for the generalized coordinate $q$, it is simple to show that the equation of motion associated with the above Lagrangian is
\eqn{
\frac{2}{3} (q-q_0) \ddddot{q} + \frac{4}{3}  \dddot{q} \dot{q} + \ddot{q}^2 = -\frac{a_0}{m}\frac{\ddot{q}}{|\ddot{q}|}\frac{\partial U}{\partial q}. \label{eq1}
}
Note  that, despite the presence of a MOND-like term, higher order derivative terms also appear. Besides, the first term on the left and right-hand sides of the above equation depend on the generalized coordinate difference $(q-q_0)$ [recall that $U(q)\equiv U(q-q_0)$]. This is physically expected because the origin of the coordinate system could be any and would not affect the dynamics of a physical system in any way. This could be seen more easily if one performs the coordinate change ${\cal Q}= q-q_0$, and then Eq. \eqref{eq1} could be written as,
\eqn{
\frac{2}{3} {\cal Q} \ddddot{{\cal Q}} + \frac{4}{3}  \dddot{{\cal Q}} \dot{{\cal Q}} + \ddot{{\cal Q}}^2 = -\frac{a_0}{m}\frac{\ddot{Q}}{|\ddot{Q}|}\frac{\partial}{\partial {\cal Q}} U ({\cal Q}), \label{eq_new_variable}
}
which is independent of $q_0$. 
From the solution to this equation, one sees that $q(t)=q_0+ {\cal Q}(t)$ for any $q_0$. 
Eq. \eqref{eq_new_variable} also tells us that when $t\rightarrow -t$, the equation of motion does not change. Hence, time reversal symmetry is present in the theory. Naturally, if $t\rightarrow t+t_0$, where $t_0$ is a reference time, Eq. \eqref{eq_new_variable} keeps the same. This simply shows that time is homogeneous in our description. Issues regarding space homogeneity and isotropy are less trivial and we elaborate on them in our discussion section.

An interesting and simple case to be investigated would be a Hooke-like potential $U(q)= k (q-q_0)^2/2$, where $k$ is a positive constant. 
One expects a periodic motion because the equation of motion is also symmetric under $\mathcal{Q} \rightarrow  -\mathcal{Q}$, but not necessarily harmonic.
For a restoring force and from our previous sign analysis, it is clear that $\ddot{{\cal Q}}<0$ when ${\cal Q}>0$. The symmetry of the force guarantees that it suffices investigating just the case ${\cal Q}>0$ for characterizing motions. In the above setting, Eq. \eqref{eq_new_variable} simplifies to
\eqn{
\frac{2}{3} {\cal Q} \ddddot{{\cal Q}} + \frac{4}{3}  \dddot{{\cal Q}} \dot{{\cal Q}} + \ddot{{\cal Q}}^2 = \omega^2a_0{\cal Q} \label{eq_harm_oscillator},
}
where we have defined $\omega^2\doteq k/m$ just for simplicity.  In order to solve this equation numerically, we choose as initial conditions the aspects of the test particle at its turning point, namely $\dot{\cal Q}(0)=\dddot{\cal Q}(0)=\ddddot{\cal Q}(0)=0$, ${\cal Q}(0)= {\cal Q}_0$ and $\ddot{{\cal Q}}(0)\neq 0$. From Eq. \eqref{eq_harm_oscillator}, one clearly sees that $\ddot{{\cal Q}}(0)= -\omega\;\sqrt[]{a_0 {\cal Q}_0}$.  Define the dimensionless quantity $z(t)\doteq {\cal Q}(t)/{\cal Q}_0$. With this rescaled coordinate, Eq. \eqref{eq_harm_oscillator} takes the form
\eqn{
\frac{2}{3} z \ddddot{z} + \frac{4}{3}  \dddot{z} \dot{z} + \ddot{z}^2 =  \frac{\omega^2 a_0}{{\cal Q}_0}z \label{eq_harm_oscillator_PRL}
}
and our initial conditions are now $\dot{z}(0)=\dddot{z}(0)=0$, ${z}(0)= 1$ and $\ddot{z}(0)=-\omega\;\sqrt[]{a_0/{\cal Q}_0}$. Note that for this type of force the dynamics depends upon the amplitude of the system.

Just for definiteness and motivated by STEP (see, e.g., \citep{2016PhRvL.117g1103P} and references therein), we choose $\omega=2\pi/1000\approx 6.28\times 10^{-3}$s$^{-1}$. We recall that we are in the deep MOND regime, and that $|\ddot{{\cal Q}}|$ is always smaller than $\omega (a_0{\cal Q}_0)^{1/2}$. Therefore, for a given $\omega$, just some amplitudes are physically meaningful. Table \ref{ta1} shows the periods of oscillation for some ${\cal Q}_0$ when our MOND-like dynamics  ($P_{M-l}$), Eq. \eqref{eq_harm_oscillator_PRL}, and Milgrom's dynamics ($P_{M}$) are taken into account. In both cases one notes that the periods are very different from each other and their Newtonian counterparts, which is 1000s. Therefore, experiments with oscillators in orbit could easily distinguish the dynamics if the SEP holds approximately. 
\begin{table}
\centering
\begin{tabular}{@{}c| c| c@{}} 
${\cal Q}_0$(nm) & $P_{M-l}$(s) & $P_{M}$(s) \\ \hline
0.01 & 39.11 & 42.39 \\
0.1 & 69.55 & 75.38 \\
1.0 & 123.68 & 134.03 \\
10.0 & 219.95 & 238.36\\
100 & 391.12 & 423.87 \\
\end{tabular} 
\caption{Oscillation period (in seconds) for a Hooke law potential using the pure deep MOND ($P_M$) equation of motion and for our MOND-like equations of motion ($P_{gM}$). In both cases the period depends on the stretched length $\mathcal{Q}_0$.}\label{ta1}
\end{table}

Since our toy model has higher order terms in time derivatives, a couple of words about the Ostrogradsky's instabilities \cite{Ostrogradsky:1850fid,Woodard:2015zca} are in order. 
The periodic oscillator above is not \textit{non-degenerate} at the origin, thus leading us to have a piecewise Lagrangian. This piecewise formulation avoids Ostrogradsky's theorem and it could have a stable solution even with a higher derivative differential equation of motion. Indeed, this is qualitatively similar to piecewise Lagrangians in the context of nonlinear eletrodynamics, where no-go theorems regarding regular black holes with electric charges can also be circumvented \cite{2001PhRvD..63d4005B}. 

\section{MOND-like dynamics for central potentials}\label{MOND3D}

We start with the three-dimensional case and then restrict our analysis to the 2D case in order to fix the parameters in our MOND-like Lagrangian. Spherical coordinates are chosen here, as they are natural to central forces such as gravity. As motivated by standard MOND, we seek for a Lagrangian  parameterized by the position and acceleration of a test particle, instead of its position and velocity. The orbits of test particles will be of special interest for determining $L$ due to the known properties of rotation curves, and for simplicity we will specialize our analysis to the case of circular ones. After having extracted the parameters of the Lagrangian, one could immediately write it in terms of generalized coordinates and apply it to the description of systems under any other interactions.

The acceleration vector in spherical coordinates can be written in terms of the coordinates $(r,\theta,\phi)$ as ($\theta$ is the polar angle with respect to the north pole and $\phi$ the azimuthal angle)
\begin{eqnarray}
\vec{a} &=& (\ddot{r}-r\dot{\phi}^2\sin^2\theta-r\dot{\theta}^2)\hat{e}_{r}
+ (r\ddot{\theta}+2\dot{r}\dot{\theta} - r \dot{\phi}^2\sin\theta\cos\theta)\hat{e}_{\theta}\nonumber\\
&+& (r\ddot{\phi}\sin\theta+2\dot{r}\dot{\phi}\sin\theta + 2r\dot{\phi}\dot{\theta}\cos\theta)\hat{e}_{\phi}\label{acceleration_spherical}.
\end{eqnarray}
From our Newtonian knowledge and the 1D deep MOND case analyzed in Sec.~\ref{1Dmond_setup}, it is clear that the 3-D version of our MOND-like Lagrangian should be
\begin{equation}
L=-\frac{m}{2}\beta\left(\frac{|\vec{a}|}{a_0}\right)\vec{a}\cdot\vec{r}- U(r)\label{3DLagrangian},
\end{equation}
where $\beta(x)$ is the generalization of the interpolating function and should be the unit when $|\vec{a}|\gg a_0$ or when $x\gg 1$. Besides, $m$ is the inertial mass of a test particle and will be assumed to be constant and $U(r)$ is a central potential, associated with a central force via the usual relation $\vec{F}=-\nabla U=-(\partial U/\partial r)\, \hat{e}_r$.

Let us see whether or not the equations of motion associated with Eq. (\ref{3DLagrangian}) admit the solution $\theta = \pi/2$. We expect this to be the case given that it happens in the Newtonian case. The equation of motion for $\theta$ is given by
\begin{equation}
-\frac{d^2}{dt^2}\left(\frac{\partial L}{\partial \ddot{\theta}} \right) +\frac{d}{dt}\left(\frac{\partial L}{\partial \dot{\theta}} \right) - \frac{\partial L}{\partial \theta}=0\label{theta_eq}.
\end{equation}
For future convenience we define
\begin{equation}
a\doteq |\vec{a}|= \sqrt[]{a_r^2+a_{\theta}^2+ a_{\phi}^2}\label{accel_magnitude},
\end{equation}
where $a_i$ is the i$th$ component of $\vec{a}$ in the i$th$ direction. 

From the above definition, it does follows that $\partial a/\partial \ddot{\theta}$ is null for $\theta=\pi/2$. Analogously, $\partial a/\partial \theta = \partial a/\partial \dot\theta = 0$ for $\theta=\pi/2$. Besides, $\partial a_r/\partial \dot{\theta}=\partial a_r/\partial \theta=0$ for $\theta=\pi/2$. Gathering these results, $\partial L/\partial \dot{\theta}=\partial L/\partial \theta = \partial L/\partial \ddot{\theta} = 0$ for $\theta=\pi/2$. 
From all the above we conclude that Eq.~(\ref{theta_eq}) is identically satisfied, which means indeed $\theta=\pi/2$ is solution to the MOND-like equations and hence motions solely in the equatorial plane exist in our local modified inertia dynamics.

Hereafter we restrict ourselves to the case of equatorial motions, motivated by the kinematics of spiral galaxies. In this case, Eq. (\ref{acceleration_spherical}) simplifies to
\begin{equation}
\vec{a} = (\ddot{r}-r\dot{\phi}^2)\hat{e}_r + (r\ddot{\phi}+ 2\dot{r}\dot{\phi})\hat{e}_{\phi}\label{accel_equator}
\end{equation}

Let us investigate now the properties of $\dot{\phi}$. Given that $\phi$'s equation of motion is 
\begin{equation}
-\frac{d^2}{dt^2}\left(\frac{\partial L}{\partial \ddot{\phi}} \right) +\frac{d}{dt}\left(\frac{\partial L}{\partial \dot{\phi}} \right) - \frac{\partial L}{\partial \phi}=0\label{phi_eq},
\end{equation}
it follows from Eq. \eqref{3DLagrangian} that 
\begin{equation}
-\frac{d}{dt}\left(\frac{\partial L}{\partial \ddot\phi} \right)+ \frac{\partial L}{\partial \dot{\phi}}= \mbox{const}\label{first_integral_phi}.
\end{equation}
One notes here that $\phi$'s dynamics is much more complicated in the MOND-like case than in Newton's dynamics. First of all, in Newton's case $\partial L/\partial \ddot \phi=0$, while this is generally not the case for our MOND-like dynamics, due to $\beta (a/a_0)$. However, things get simpler in the case $r=$const., which reflects circular motions, as we study now. 

When orbits are circular in our MOND-like's dynamics and one takes the ansatz $\dot{\phi} = $const., it follows that $\partial a/\partial \ddot{\phi}=0$. Thus, $\partial L/\partial \ddot{\phi}= 0$. Besides, $\partial a/\partial \dot{\phi}= $const., which implies that $\partial L/\partial \dot{\phi}=$const. 
Therefore, Eq. (\ref{first_integral_phi}) is satisfied identically, meaning that when $r=$const., $\dot{\phi}$ is
also a constant. It is exactly what one expects because in a circular orbit all of its points are equivalent and hence $\dot{\phi}$ should be the same for all of them.

We are now in the position of assessing the consistency of our approach (taking into account a local Lagrangian with a modified acceleration term) with flat rotation curves. From $r=$ const. and $\dot{\phi}=$ const., one obtains that $\partial L/\partial \ddot{r}= $const., $\partial L/\partial \dot{r}=0$, which leaves the equation of motion of $r$ to be
\begin{equation}
\frac{\partial L}{\partial r}=0\label{eq_of_motion_circular}.
\end{equation}
Thus, from Eqs. (\ref{3DLagrangian}) and (\ref{accel_equator}),
\begin{equation}
\frac{m}{2}\frac{\partial \beta}{\partial a}\frac{\partial a}{\partial r}r^2\dot{\phi}^2+ m\beta r\dot{\phi}^2=\frac{\partial U}{\partial r}.
\end{equation}
In this particular case one can easily show that $\partial a/\partial r= \dot{\phi}^2$. From $\dot{\phi}= v/r$, with $v$ the magnitude of the tangential velocity, and the above equation, one thus has
\begin{equation}
\frac{m}{2}\frac{\partial \beta}{\partial a}\frac{v^4}{r^2}+ m\beta \frac{v^2}{r}= \frac{\partial U}{\partial r}.
\end{equation}
If one now takes $\beta =c_1a/a_0$ ($c_1$ is a constant) and specializes to the gravitational case $U=-MGm/r$ (M is the total baryonic mass of a galaxy, which we take as a constant), it finally follows that [$a=r\dot{\phi}^2$ for circular orbits, as given by Eq. (\ref{accel_equator})]
\begin{equation}
\frac{3}{2}c_1v^4=MGa_0\label{Tully_Fisher}.
\end{equation}
By taking $c_1=2/3$ one gets the asymptotic constancy of the tangential velocity of test particles in circular orbits in spiral galaxies, exactly as in standard MOND.

In summary, our local analysis with a modified acceleration term in the Lagrangian also reproduces, in the case of circular orbits, the results of standard Milgrom's dynamics. Therefore, Milgrom's prescription to a modified dynamics from rotation curves is not unique, and there are local theories for modified inertia which could raise from them.

\subsection{Conserved charge - deep MOND-like case} 

As we pointed out previously, for a circular motion, the Lagrangian
\begin{equation}
L=-\frac{m}{3} \frac{|\vec{a}|}{a_0}\vec{a}\cdot\vec{r}- U(r),
\label{model1}
\end{equation}
yields the following equation of motion in the radial direction:
\begin{equation}
\frac{m}{a_0}r^2\dot{\phi}^4= m \frac{|\vec{a}|^2}{a_0} = 
\frac{\partial U}{\partial r}.\label{radial_eom_deep_mond}
\end{equation}
which is the expected deep MOND regime and is well known to recover the Tully-Fisher relation if one considers Newton's gravitational force on the RHS of this equation \cite{Milgrom:1983ca}. Note, however, that in this set up Eq.~\eqref{radial_eom_deep_mond}, or $L$ given by \eqref{model1}, is more general than that since it is valid for any potential $U(r)$.

Lagrangian (\ref{model1}) does not depend explicitly on $\phi$, so the generalized Euler-Lagrange equation for this component provides the following constant of motion
\eqn{
Q_{\phi} &=& -\frac{d}{dt}\left(\frac{\partial L}{\partial \ddot\phi} \right)+ \frac{\partial L}{\partial \dot{\phi}}= \frac{4m}{3a_0} \dot{\phi}^3 r^3,\label{genangular}
}
which is the Noether charge in the deep MOND regime, and for the second line of the above equation circular orbits have been selected. Note that it is different from the usual angular momentum of Newtonian mechanics and it is not only valid for the gravitational interaction. It is trivial however to check that it is valid for the gravitational interaction since $v = \dot{\phi}r$ is known to be constant by the Tully-Fisher relation. 

\subsection{Conserved charge - general case}

The conserved charge, for circular motions, associated to the Lagrangian (\ref{3DLagrangian}) reads
\eqn{
Q_{\phi} &=& m\dot{\phi} r^2 \bigg(\beta(|\vec{a}|/a_0) + |\vec{a}| \frac{\partial \beta}{\partial |\vec{a}|}\bigg). \label{genangular2}
}
Note that for $\beta = 1$, the Noether charge is the Newtonian angular momentum and for $\beta = \frac{2}{3} \frac{|\vec{a}|}{a_0} $ it gives (\ref{genangular}) as it should be. 

\subsection{Modified area law }

A different conserved quantity will modify Kepler's second law. 
The variation with respect to time of an area orbit spanned by a radius vector $r(t)$ is easily computed to give
\eqn{
\frac{dA}{dt} = \frac{1}{2}r^2 \dot{\phi}.
}
Rewriting $\dot{\phi}$ in terms of the constant of motion given by (\ref{genangular}) yields for circular orbits
\eqn{
\frac{dA}{dt} = \bigg(\frac{3Q_\phi a_0}{32m}\bigg)^{1/3}
r,}
\textit{i.e.}, in the deep MOND regime the area speed is directly proportional to the distance of the test particle to the center of force. Note that this is constant for a circular motion, but it is sensitive to perturbations in the radial direction. This is a prediction that could be tested in future experiments using, for example, electromagnetic interaction of particles in an almost circular motion. 

\section{Perturbations in 2D MOND-like}\label{MOND2D}

After having rederived successfully the results of MOND in the particular case of a circular orbit by using a local Lagrangian, it is important to study whether this solution is stable under small perturbations, and, if not, what are the characteristic instability timescales. We now move on to investigate that.

\subsection{Deep MOND regime}

As a toy model, we consider a test particle in a plane in deep MOND regime subjected to a central force which can be described by the simple ansatz given by the Lagrangian (\ref{model1}). Considering its initial motion as circular at $r(t)=R$, one has that when small perturbations of the orbit are present 
\eqn{
r(t) &=& R + \delta r(t),
\\
\dot{\phi}(t) &=& \dot{\phi}_0 + \delta \omega(t),
\\
\theta(t) &=& \frac{\pi}{2}+ \delta \theta(t),
}
where $\delta \theta(t) = \delta z(t)/R$ are small perturbations in the vertical direction ($z$).
Generalized Euler-Lagrange equations yield at first order in $\delta r$, $ \delta \omega$ and $\delta \theta$ the following coupled system of differential equations:
\eqn{
-12 \dot{\phi}_0^3R^2 \delta \omega +3\bigg(a_0 V''(R)- 2\dot{\phi}_0^4R\bigg)\delta r + 2R\left(3\dot{\phi}_0R \frac{d^2 }{dt^2} \delta \omega + 6 \dot{\phi}_0^2\frac{d^2 }{dt^2}\delta r- \frac{d^4 }{dt^4}\delta r   \right)=0,\label{pert_circ_1} 
}
\eqn{
12\dot{\phi}_0^2R\frac{d}{dt} \delta \omega + 12\dot{\phi}_0^3\frac{d}{dt}\delta r -R \frac{d^3}{dt^3} \delta \omega - 6 \dot{\phi}_0\frac{d^3}{dt^3}\delta r=0,
\label{pert_circ_2} 
}
\eqn{
\frac{d^4}{dt^4}\delta \theta -2 \dot{\phi}_0^2 \frac{d^2}{dt^2}\delta \theta - 3 \dot{\phi}_0^4 \delta \theta=0,
\label{pert_circ_3} 
}
where, just for simplicity, we have defined $V(R)\doteq U(R)/m$. Note that, at first order in perturbations, the vertical component does not couple to the planar coordinates. We see that (\ref{pert_circ_2}) is a total derivative. Simple algebraic manipulations lead to
\eqn{
\delta \omega = \frac{1}{60 \dot{\phi}^3_0 R^2} \bigg(-3 \bigg[a_0 V''(R)+22\dot{\phi}^4_0 R\bigg]\delta r +2R\bigg[3 A \dot{\phi}_0+12\dot{\phi}_0^2 \frac{d^2}{dt^2}\delta r+\frac{d^4}{dt^4}\delta r\bigg]\bigg)\label{omega}
}
where $A$ is a constant equal to the total derivative term of (\ref{pert_circ_2}). Plugging Eq. \eqref{omega} into (\ref{pert_circ_1}) we get an equation for $\delta r$ only. Using $V''(R)= - 2GM/R^3$,
$v^4=(\dot{\phi}_0 R)^4=MGa_0$ (Tully-Fisher relation), it follows that the master equation for $\delta r$ is simply given by
\eqn{
R^4\frac{d^6}{dt^6}\delta r+ 6 MGa_0\frac{d^2}{dt^2}\delta r - 6 A (MGa_0)^{\frac{3}{4}}R=0 \label{eq_deep_MOND_simplified},
}
and its general solution is 
\eqn{
\delta r(t) = c_1 + c_2 t + c_3 t^2 + \sum_{n} A_n\Re\left(e^{i\Omega_n t}\right),
\label{radial}}
where $c_1$, $c_2$ and $A_n$ are real constants that can be fixed by initial conditions, $c_3 = A R/2(a_0 GM)^{1/4}$ and $\Omega_n=6^{\frac{1}{4}}(v/R)e^{i\frac{n\pi}{4}}$, $n=1,3,5,7$. 
Note that $\Omega_n$ are complex numbers, which implies that harmonic solutions to $\delta r$ are modulated by exponentially increasing and decreasing functions. 
The solution for the angular velocity is then
\begin{eqnarray}\label{delta_omega}
&&\delta \omega (t) = d_1 + d_2 t + d_3 t^2 - \frac{1}{30(a_0GM)^{\frac{3}{4}}}\sum_nA_n\times\\ &&\Re\left( e^{i\Omega_nt}[30a_0GM+12(a_0GM)^{\frac{1}{2}}R^2\Omega_n^2 - R^4\Omega_n^4]\right),\nonumber
\end{eqnarray}
where 
\eqn{
d_1 &=& A \frac{R}{2\sqrt{a_0 GM}} - c_1 \frac{(a_0 G M)^{1/4}}{R^2},\\
\label{d2}
d_2 &=& - c_2 \frac{(a_0 G M)^{1/4}}{R^2} \\
d_3 &=& - \frac{A}{2R}.
}
The solution to the vertical perturbation is
\eqn{
\delta \theta(t) = b_1 \cos(\dot{\phi}_0 t) + b_2 \sin(\dot{\phi}_0 t) + b_3 e^{-\sqrt{3}\dot{\phi}_0 t} + b_4 e^{\sqrt{3}\dot{\phi}_0 t}\nn.\\
}

\subsubsection{Instabilities}

We see that all solutions are unstable in general. That is in fact already expected by the Ostrogradsky theorem since our 2-D Lagrangian is non-degenerate everywhere in the phase space. Circular orbits avoid Ostrogradsky theorem by providing equations of motion that naturally excludes higher order derivative terms.

From the simple fact that even in the limit where there are no perturbations of circular orbits ($\delta \omega$, $\delta r$ and $\delta \theta$ are null) Eq. \eqref{omega} should be satisfied, it follows that $A$ in the above equations is null. This implies that $c_3 = d_3 = 0$ and instabilities that are quadratic in time are not present.
Also, note that only if $A_n=0$ the solutions given by Eqs. \eqref{radial} and \eqref{delta_omega} automatically imply $\delta v=\delta(r\dot{\phi})=0$ at all times (since $A=0$). Then, for $A_n\neq 0$, the Tully-Fisher relation is violated in our approach. This would imply an experimental fluctuation of $v$ in the outskirts of galaxies. Since there is no observational evidence for such scattering \cite{Lelli:2015wst}, one can set $A_n = 0$\footnote{This is similar to what one does when solving the wave equation in quantum mechanics. However, timescales for fluctuations on regions far enough from the center of galaxies could be larger than the age of the universe, and hence they should be negligible even when $A_n$'s are nonzero. This will be discussed later on.}. A similar argument follows for the motion in the vertical direction. The exponential divergent solution would introduce a large scattering in the observations. Since it is not present one must set it to zero, \textit{i.e.}, $b_4=0$. The constant $c_1$ in Eq. \eqref{radial} can always be taken as null if one assumes particles are perturbed from the initial circular orbit, and such a choice implies $d_1 = 0$. Taking into account all the conditions above, the solutions for the perturbations are:
\eqn{
\delta r(t) &=&  c_2 t
\\ \label{theta}
\delta \omega (t) &=& d_2 t,
\\ \label{vert}
\delta \theta(t) &=& b_1 \cos(\dot{\phi}_0 t) + b_2 \sin(\dot{\phi}_0 t) + b_3 e^{-\sqrt{3}\dot{\phi}_0 t}.
}
Note that the planar solutions are still unstable but they do not increase the acceleration, leaving the system in the deep MOND regime. This is a consistency check, since the restrictions imposed on the general solutions were based in the deep-MOND regime, which is now preserved.

Take as the typical velocity of instabilities the dispersion velocity in spiral galaxies, $v_r$. Then, $c_2\approx v_r$. Perturbative analysis breaks down when $\delta r\approx R$. Hence, $\tau_{c_2}\approx R/v_r$. In the deep MOND region, one assumes $a\ll a_0$. For practical ends, we will take as deep MOND $a\lesssim 10^{-1}a_0$. In terms of characteristic distances, this means $R\gtrsim 10 v_\varphi^2/a_0$. If one takes $v_\varphi=100$ km s$^{-1}$ as a representative orbital velocity, it follows that $R\gtrsim 10^{23}$cm $\approx 30$ kpc. The typical error associated to the measurement of $v_\varphi$ is less than $10\%$ of $v_\varphi$ \cite{Li:2018tdo}, thus an upper bound for the radial velocity is $v_r = v_\phi/10$. Assuming $v_r \lesssim  v_\varphi/10$, it implies that the perturbation theory breaks down at $\tau_{c_2} \gtrsim  3$ billion years. This is the most conservative estimate because we are overestimating radial velocities. More realistic estimates could have much larger timescales. If $v_r$ is around 1\% of $v_{\phi}$ (around $10\%$ of the total scattering of the orbital velocity), which is a fair estimate as well, one reaches over 30 billion years for the timescales. Hence, the range of acceptable values to the timescales is very broad and it could be easily larger than the age of the universe. As one can see from Eqs. \eqref{theta} and \eqref{d2}, perturbation theory for the tangential velocity $v_\phi$ breaks down when $\tau_{d_2} \approx  R/v_r$, which is the same as $\tau_{c_2}$.

\subsubsection{Vertical perturbations}

Differently from plain MOND, our approach allows the split of the vertical and planar motions, similarly to the Newtonian case. More than that, it has recently been argued that the vertical motion of stars in the outskirts of spiral galaxies favors a dark matter (DM) halo over a MOND description \cite{Stacy:2015a,Lisanti:2018qam}, while the small scattering observed in the orbital velocity dispersion favors exactly the opposite \cite{Lelli:2015wst}. From Eq. \eqref{vert}, one can see that after a timescale of order $\tau \approx 1/\dot{\phi}_0 \approx 300$ million years, the solution to $\delta \theta$ quickly converges to the {\it same} solution from a DM halo. Therefore, galaxies older than $\tau$ should not behave any differently, in the vertical direction, than the motion predicted by Newtonian theory with dark matter. However, deviations from the DM description in young galaxies (aged less than 300 Myr) could distinguish our proposal from the DM scenario when $b_3 \neq 0$.\footnote{The absence of any deviations from the Newtonian description with DM can also be used to set $b_3= 0$.}

The set up proposed here seems to unify the best of both scenarios. It yields the baryonic Tully-Fisher relation with negligible scattering in the planar motion and the oscillatory behavior, characteristic of the Newtonian theory, in the vertical one.

\subsection{Intermediate MOND regime}

Here we investigate perturbations of circular orbits with a particular choice for the usual interpolating function $\mu$. We work with $\mu$ found in Ref. \citep{2016PhRvL.117t1101M} from analysis over two thousand data points in more than a hundred spiral galaxies with very different aspects. It is
\begin{equation}
    \mu(a)=1-e^{-\frac{a}{a_0}}\label{mu_stacy}.
\end{equation}
Clearly from this equation, when $a\gg a_0$, one recovers the Newtonian limit and when $a\ll a_0$, deep MOND regime is obtained. Eq.~\eqref{mu_stacy} implies that both Newtonian and MONDian limits are reached quickly, meaning intermediate MOND regimes are more fleeting. However, these regimes are also observed in rotation curves, which naturally render them relevant. This is specially so in our analysis, given that in deep MOND regime perturbations are unstable. One expects to also find unstable solutions, given that Ostrogradsky's criteria are not violated in this gravitational case, and the relevant point is the timescale instabilities. 

From deep MOND analysis in our approach, one learns that $\beta(\frac{a}{a_0})=\mu(\frac{2a}{3a_0})$, and it follows from Eqs.~\eqref{3DLagrangian} and \eqref{mu_stacy} that the relevant Lagrangian now is 

\begin{equation}
    L=-\frac{m}{2}\left( 1-e^{-\frac{2a}{3a_0}} \right)\vec{a}\cdot \vec{r}-U(r)\label{Lagrangian-stacy-mu}.
\end{equation}
Perturbation analysis follows the same procedure of the last section and here we just present the main results. When $U(r)=-GMm/r$ is taken, one has that the perturbed equations for $\delta r$ and $\delta \omega$, using $\dot{\phi}_0 = v/R$ for the background circular orbit, can be written as
\begin{eqnarray}\label{interm1}
C_1 \delta r &+& C_2  \frac{d^2}{dt^2} \delta r + C_3 \frac{d^4}{dt^4} \delta r + C_4 \frac{d^6}{dt^6} \delta r =0,
      \\\label{interm2}
    B_1 \delta \omega &=&  B_2 \delta r
        +B_3\frac{d^2}{dt^2} \delta r +
      B_4 \frac{d^4}{dt^4} \delta r,
\end{eqnarray}
where
\begin{eqnarray}
C_1 &=& 6 \bigg(6 a_0^3 G^2 M^2 R^3 v^2 +
      12 a_0^2 G M R^2 v^4 (G M - R v^2) +
      v^8 (3 G^2 M^2 + G M R v^2 - 2 R^2 v^4)
      \nonumber\\
      &+&      a_0 R v^6 (-9 G^2 M^2 - 2 G M R v^2 + 6 R^2 v^4)\bigg),
      \\
      C_2 &=&  R^2 \bigg(36 a_0^3 G^2 M^2 R^3 - 45 G^2 M^2 v^6 + 72 G M R v^8 - 
      28 R^2 v^{10} + 18 a_0^2 G M R^2 v^2 (G M - 2 R v^2)
      \nonumber\\
      &+&
      6 a_0 R v^4 (9 G^2 M^2 - 17 G M R v^2 + 
         8 R^2 v^4)\bigg),
   \\
   C_3 &=& 
   2 R^4 (3 G M - 
      2 R v^2) \bigg(9 a_0^2 G M R^2 + 6 a_0 G M R v^2 - 6 G M v^4 - 
         9 a_0 R^2 v^4 + 5 R v^6 \bigg),
         \\
C_4 &=&  R^6 (3 G M - 
      2 R v^2)^2 (2 a_0 R - v^2) ,
      \\
      B_1 &=& 2 R^2 v (2 a_0 R - 
     v^2) (2 a_0^2 (-1 + e^{v^2/(a_0 R)}) R^2 + 13 a_0 R v^2 - 4 v^4),
     \\ 
     B_2 &=& 4 a_0^2 (-13 + 2 e^{v^2/(a_0 R)}) R^2 v^4 + 27 a_0 R v^6 - 4 v^8 - 
      2 a_0^3 R^2 (-11 R v^2 +\nonumber\\ 
        &+& e^{v^2/(a_0 R)} (-3 G M + 11 R v^2)),
    \\
    B_3 &=& 
   R^2 \bigg(-2 a_0^3 (-1 + e^{v^2/(a_0 R)}) R^3 + 26 a_0^2 R^2 v^2 - 22 a_0 R v^4 + 4 v^6\bigg) ,
\\
 B_4 &=&    a_0 R^5 (2 a_0 R - v^2).
\end{eqnarray}

Equations \eqref{interm1} and \eqref{interm2} can be easily solved analytically for spiral galaxies and one learns that all solutions should have the form $e^{\Omega t}$, with $\Omega$ an imaginary constant, fixed by the parameters. As a convention, we will assume that the intermediate MOND region of a galaxy is characterized by $10^{-1}a_0 \lesssim a \lesssim 10a_0$. In terms of radii from the galactic center, this roughly translates into $1/10 (v^2/a_0)\lesssim R\lesssim 10(v^2/a_0).$ For instance, take the case of spiral galaxies with $v$ of the order of $100$~km~s$^{-1}$. This roughly implies that $10^{21}\lesssim R/cm\lesssim 10^{23}$ ($1 \lesssim R/kpc\lesssim 30$). As a reference, let us take $R=10$ kpc. 
(Examples of galaxies presenting tangential velocities of the order of $100$ km s$^{-1}$ at this distance are UGC6930, UGC6983, NGC4183, NGC3769, NGC3917, UGC6917 \citep{2002ARA&A..40..263S}.) Assume also that asymptotic orbital velocities are also of order $100$ km s$^{-1}$. From the Tully-Fisher relation, it follows that the mass of this galaxy should be $M=v^4/(G a_0)\approx 10^{43}-10^{44}$g ($\approx 10^{10}-10^{11}$ M$_{\odot}$). It turns out in this case that all six $\Omega$s have real and imaginary parts, which shows some solutions are unstable. For the above parameters, 
$\tau_{R=10kpc}\approx 10\times (\Im{\Omega}_{R=10 kpc})^{-1}\approx 4 \times 10^{16}$s $\approx 900$ million years. This instability timescale is not particular to the interpolating function given by Eq. \eqref{mu_stacy}. We have checked that others, such as the ``n-family''\footnote{Note however that the ``n-family'' here is implemented in the Lagrangian, not in Milgrom's equation of motion.} \citep{2016PhRvL.117g1103P}, also give very similar timescales. 

We stress that for the intermediate regions (intermediate accelerations) of spiral galaxies, exponentially growing perturbations should not be discarded in principle. Just observations -- specially scattering -- would tell if this is the case. Given the above instability timescales in these regions, if exponentially blowing up solutions to perturbations are possible, they are also very interesting for constraining our approach. However, one should also bear in mind that runaway solutions, if really present, should be ``fleeting'' in MOND-like approaches. The reason is due to their Newtonian saturation when accelerations are large, which is Ostrogradsky instability free. This means that if such instabilities could set in, which would imply in acceleration growth in general, the dynamics itself would naturally make them die away. This ``self-regulatory'' mechanism is intrinsic to MOND approaches with higher order derivatives and actually shows that Ostrogradsky instabilities could be a very subtle issue there. Further details about this are well beyond the scope of this work, and we leave them to be investigated elsewhere.

\section{Center of mass motions in the MOND-like description}
\label{CM}
For the case of our dynamics, it is important to start addressing the issue of the motion of the center of mass (CoM) of a composite system, as well as its internal dynamics. The reason is because our approach is nonlinear and hence the decoupling from internal and external dynamics might not be trivial. In the previous sections we have assumed that the motion of the CoM of a test particle under small accelerations is oblivious to aspects of its internal dynamics. This has allowed us to draw important conclusions about radial and vertical perturbations of test particles. We have found the surprising result that vertical motions, in first order perturbations, decouple from radial ones and could lead to very similar results as pertaining to Newtonian prescriptions with dark matter. For the radial perturbations, it is always possible to find situations with very small scattering of the Tully-Fisher relation. Since these results are in total agreement with observations and derive from the above mentioned assumption, we should discuss which are the conditions in our description that lead to its verification.
 
Clearly, this issue in general is nontrivial, and here we just sketch a possible way of dealing with it for our local Lagrangian. In the context of Milgrom's nonlocal description, it is already known the solution for the decoupling of internal and center of mass motions, and that relies exactly on nonlocality \citep{1999PhLA..253..273M,1994AnPhy.229..384M}. In order to have insights into some consequences of changes of inertia in local theories, we limit our analysis here to the problem of two bodies.

Based on our Lagrangian  \eqref{3DLagrangian}, we assume that the Lagrangian of a two body system under an external field $U_e$ is given by
\begin{equation}
    L=-\sum_{i=1}^2\frac{m_i}{2}\beta\left(\frac{|\ddot{\vec{r}}_i|}{a_0}\right)\ddot{\vec{r}}_i\cdot\vec{r}_i- U(\vec{r})-\sum_{i=1}^2 U_e(\vec{r}_i)\label{two_body_lagrangian}
\end{equation}
where $\vec{r}_i$ is the position vector of the $i$th particle with respect to an inertial frame, $\vec{r}\doteq \vec{r}_2-\vec{r}_1$ (the usual relative coordinate) and $U$ is the interaction potential between the two particles. We take the CoM of this system, $\vec{R}$, as given by its normal definition \citep{1969mech.book.....L}. This allows us to write
\begin{eqnarray}
     \vec{r}_1&=& \vec{R}-\frac{m_2}{M}\vec{r},\;\;
     \vec{r}_2= \vec{R}+\frac{m_1}{M}\vec{r}\label{coordinate_change},
\end{eqnarray}
with $M\doteq m_1+m_2$. Let us assume in what follows that the internal acceleration is high [$|\ddot{\vec{r}}|\gg (|\ddot{\vec{R}}|,a_0)$] and that the mass of a particle, say, $m_2$, is much larger than the other ($m_2\gg m_1$).
Thus, we will work in the limit $m_1/m_2\rightarrow 0$.
In this case, from Eq. \eqref{coordinate_change}, it follows that $\vec{r}_2\approx \vec{R}$ ($\ddot{\vec{r}}_2\approx \ddot{\vec{R}}$), $\vec{r}_1\approx \vec{R}-\vec{r}$ ($\ddot{\vec{r}}_1\approx -\ddot{\vec{r}}$). Thus, the Lagrangian \eqref{two_body_lagrangian} can be approximated by
\begin{eqnarray}
    L&\approx&-\frac{M}{2}\beta\left(\frac{|\ddot{\vec{R}}|}{a_0}\right)\ddot{\vec{R}}\cdot\vec{R}- U_e(\vec{R})-\frac{m_1}{2}\ddot{\vec{r}}\cdot\vec{r} +\frac{m_1}{2}\ddot{\vec{r}}\cdot \vec{R}- U(\vec{r}),\label{two_body_lagrangian_approximate}
\end{eqnarray}
where we have approximated $\vec{r}_i$ by $\vec{R}$ in the external potential and $U_e\doteq \sum_i U_e^i(\vec{R})$ is the total potential at the CoM. From Eq. \eqref{two_body_lagrangian_approximate}, it is easy to see that in the limit $m_1/m_2\rightarrow 0$, the influence of the crossed term $m_1\ddot{\vec{r}}\cdot \vec{R}/2$ is negligible for both the internal ($\vec{r}$) and external ($\vec{R}$) equations of motion. Therefore, in this limit, they decouple. Besides, we find a Newtonian description for the internal dynamics when internal accelerations are large and, concomitantly, a MOND-like dynamics for the center of mass.

We stress that these points are far from complete and deeper studies should also be carried out regarding other scenarios and their consequences. Notwithstanding, we leave them to be done elsewhere. The important message, up to this point, is the possibility of decoupling internal and external motions in our dynamics for convenient limits, in the same qualitative spirit as usual modified gravity MOND \cite{Bekenstein:1984tv}.

\section{Discussions and Conclusions}
\label{conclusions}

We have proposed a local Lagrangian for MOND as modified inertia whose equations of motion present higher order derivatives for a point particle in an arbitrary potential. This local Lagrangian modifies the inertial properties of such point particle leading to, at small accelerations, a dynamics that is different from the Newtonian one. For Hooke-like forces, our approach leads to stable solutions even in the presence of higher order derivatives, thus suggesting piecewise Lagrangians could be a way to get around Ostrogradsky instabilities. In the case of a gravitational central force, for a circular motion, we showed that the point particle equations of motion is in general the one described by MOND. We have also shown that in the deep MOND regime, the motion is unstable under small radial and vertical perturbations and has characteristic timescales larger than $3$ billion years. Although an upper limit to instability timescales is not definitely known, reasonable estimates point to values larger than the age of the universe.
A local Lagrangian that interpolates the Newtonian and the deep MOND regimes also leads  to unstable solutions under small perturbations in general, and timescale instabilities would be smaller than 1 billion years. We discuss the above results in light of observables now.

In the deep-MOND regime, we find unstable perturbative solutions around circular orbits due to the higher order derivative terms. These solutions lead to power-law and exponential instabilities, as expected by the Ostrogradsky theorem.  The latter are characterized by the coefficients $A_n$, which lead to a scattering of the Tully-Fisher relation. Since such scattering is not observed \cite{Lelli:2015wst}, these coefficients should be all set to zero phenomenologically. The remaining instabilities provide a time frame for which the perturbation theory remains valid. The most conservative estimate gives us a time frame of $3$ billion years, after which perturbation theory breaks down. It is important to emphasize that our dynamics naturally introduces an asymptotic kinematic limit, the one of very large accelerations (the ultraviolet limit in natural units), in which the Newtonian dynamics is recovered effectively, and Ostrogradsky instabilities are absent. Thus, the linear perturbation instabilities will necessarily be tamed in a full nonperturbative analysis in the Newtonian limit, whose precise analysis is left for future work. The combination of such phenomenological and theoretical considerations may be seen as a ``physical'' way to overcome Ostrogradsky's instabilities, similarly to what is done when unstable solutions in quantum mechanical systems are discarded based on boundary conditions and solutions corresponding to quantum interference between states vanish as the classical limit is considered. Similar considerations follow for the intermediate MOND regime, where the typical timescale is a little shorter than $1$ billion years . As in the deep MOND regime, just observations in the intermediate region will tell if runaway solutions to perturbations should be present in our description. If the scattering is large, then this is an indication that instabilities should take place. However, as mentioned before, they should be ``fleeting'' because the dynamics would eventually converge to Newton's. Thus, comparison of intermediate regions of similar mass galaxies at different redshifts might be a powerful tool for constraining our local MOND-like Lagrangian.

Moreover, we emphasize the vertical perturbations have been shown to decouple from radial motions in linear perturbations and their solutions are stable (when exponentially growing perturbations are physically eliminated in our MOND-like proposal in the deep MOND region). After a characteristic timescale of some hundred of millions of years (when exponentially damping solutions are negligible), its outcome is very similar to the ones coming from dark matter approaches. Thus, our approach could explain simultaneously the small scattering of the Tully-Fisher relation and the stability (harmonic oscillations) of vertical perturbations. That is, the proposed dynamics could unify some of the phenomenology of the usual MOND and dark matter. Concomitantly, it would naturally point to scenarios where it could be either differentiated from dark matter or falsified. For instance, young  enough  galaxies (aged less than a billion years) could present a different vertical behavior from older ones in our approach, exactly due to the exponentially damping term. This  would  be  a  prediction  of our description and  a  smoking  gun  to it.

We stress that in the purely gravitational case the acceleration does not change sign and hence only a branch of our MOND-like Lagrangian is relevant; therefore, Ostrogradsky instabilities should always set in in general, as we have shown explicitly. However, the interesting aspect of our description is that instabilities in the very deep MOND region are tackled phenomenologically due to the very small scattering of the Tully-Fisher relation together with the Newtonian limit recovered for large accelerations. When Hooke-like forces are at play, on the other hand, the acceleration of a test particle could change sign. From Eq. \eqref{L3D_simplest}, one has that the Lagrangian in this case is actually piecewise, which implies that certain derivatives thereof are not defined at some points. This violates
one of Ostrogradsky's theorem's conditions and it could lead to stable solutions even with a higher order differential equation. 
We note that some results of Sec. \ref{1Dmond_setup} for Earth-based laboratories should be seen as only indicative: we have assumed that the SEP is valid for the calculations and this issue is actually very subtle. 

Indeed, a pertinent question in our description is whether the Strong Equivalence Principle is violated. In the usual MOND's case this happens since the gravitational field equation becomes nonlinear, and hence the superposition principle is broken. Because of that, it becomes impossible in general to find a reference frame in which an external gravitational field can be cancelled out locally (by a convenient frame moving with a given constant acceleration) and hence the internal dynamics of a freely-falling system depends on its embedding aspects (for instance its spatial location or its motion such as velocity or acceleration). Although in our case the gravitational field equation would remain unchanged with regard to the linear Poisson equation, the nonlinearity of the equations of motion for test particles might also spoil the SEP. Indeed, if one makes the coordinate change ${\cal Q}\rightarrow {\cal Q}'+\alpha(t)$, it is not possible in general to choose a locally constant $\ddot{\alpha}$ such that a gravitational field in a small space region (approximately constant) could be eliminated. Note that the same happens to the Milgrom's usual equation of motion, given that it is also nonlinear in the acceleration due to the interpolating function. Therefore, in our approach EFE-like effects might also take place and hence Earth-based experiments should be analyzed carefully. Given its possible subtleties, we leave such analysis to be done elsewhere.

For obtaining all the astrophysical results of this work, we have tacitly assumed that the motion of the center of mass of a system decouples from its internal dynamics. Actually, this is nontrivial when the dynamics is nonlinear, as is our case. However, we have attempted to justify this by taking into account a toy model system with two particles. We have indeed found that when internal the acceleration of it is large and a mass is much smaller than the other, such a decoupling happens and the center of mass dynamics can be MOND-like, while the internal dynamics would be Newtonian. Thus, when our two body system is a rough model for an atom inside or around a much larger system (a star, for instance), our analysis would mean that its radiation (apart from the Doppler shift) should be the same as the radiation measured on a counterpart on Earth.
Naturally, further studies need to done regarding center of mass motions and internal dynamics, but we leave them for other works. In special, one might not expect a decoupling of dynamics for general physical situations and they could work as smoking guns to our proposal.

Our approach also has other interesting physical properties, which we discuss more deeply now. Our sole hypothesis has been the existence of a local Lagrangian for test particles in a MOND regime. This is reasonable given that so far all experimentally known interactions are described by local Lagrangians. With MOND's phenomenological aspects in a region with $a\ll a_0$, one can fix all asymptotic aspects to this Lagrangian, and for intermediate regimes one could define an interpolating-like function, as in usual MOND. This has been done in Sec. \ref{MOND3D}. One sees from Eq. \eqref{3DLagrangian} that, differently from the Newtonian Lagrangian, our MOND-like Lagrangian does not present space homogeneity and isotropy, given that it cannot be written as a pure function of $|\vec{a}|$ (or even $|\vec{v}|$). (It does present origin invariance, as explained in Sec. \ref{1Dmond_setup}, and this is necessary because one can set it where one wishes.) Therefore, it is not possible to define inertial coordinate systems such that the laws of motion remain the same for two reference frames moving relative to each other with a constant velocity. This result is expected in our approach since, as stated by Milgrom (see, e.g. \citep{Milgrom:2005mc,Milgrom:2011kx} and references therein), MOND as modified inertia fulfilling the Galilean invariance has to be nonlocal. The violation of the Galilean principle in our local MOND-like theory can be clearly seen in the $1D$ case, taking the equation of motion \eqref{eq_new_variable}. Indeed, it follows from it that the transformation ${\cal Q}\rightarrow {\cal Q}'+Vt$, $V$ being a constant, does not render the equations of motion for ${\cal Q}'$ and ${\cal Q}$ the same. However, Eq. \eqref{eq_new_variable} also shows a notable property. For free-particles, ${\cal Q}={\cal Q}_0 + {\cal Q}_1t$, ${\cal Q}_0$ and ${\cal Q}_1$ are constants, is a solution. Therefore, a free particle could move with a constant velocity in our approach. Besides, in the free case, if one makes ${\cal Q}\rightarrow {\cal Q}'+Vt$, though the equations of motion are different, a solution for ${\cal Q}'$ is also ${\cal Q}'_0 +{\cal Q}'_1t$ (${\cal Q}'_0$ and ${\cal Q}'_1$ are constants). This is important because it shows that there are regular solutions to our approach and hence they should be the ones of physical interest.

\section{Acknowlegments}
We thank Gabriel Flores-Hidalgo and James Overduin for useful comments. J.P.P. acknowledges the financial support given by Funda\c c\~ao de Amparo \`a Pesquisa do Estado de S\~ao Paulo (FAPESP) under grants No. 2015/04174-9 and 2017/21384-2, and the Polish National Science Centre under grant No. 2016/22/E/ST9/00037. The research at McGill is supported in part by funds from NSERC, from the Canada Research Chair program and from a John Templeton Foundation grant to the University of Western Ontario. G.F. is also thankful to University of Cape Town and the Yukawa Institute for Theoretical Physics for their hospitality during the period in which this work was written, and acknowledges financial support from the JSPS
Fellowship. R.C. acknowledges the financial support by the SARChI NRF grant holder. 

\bibliographystyle{JHEP}
\bibliography{References}

\end{document}